Increase of superconducting order parameter in the magnetic superconductor $Dy_{0.6}Y_{0.4}Rh_{3.85}Ru_{0.15}B_4$ in a wide magnetic field range. Is it a further candidate in *p*-wave superconductors?

L.F.Rybaltchenko[1*], E.V.Khristenko[1], L.A.Ishchenko[1], A.V.Terekhov[1,2,3], I.V.Zolochevskii[1] [‡], E.P.Khlybov[2,4], and A.J.Zaleski[3].

[1] B.I. Verkin Institute for Low Temperature Physics and Engineering, NAS of Ukraine, 47 Lenin Ave., Kharkov, 61103, Ukraine.

[2] L.F. Vereshchagin Institute for High-Pressure Physics, RAS, Troitsk, 142190, Russia.

[3] W. Trzebiatowski Institute of Low Temperature and Structure Research, Polish Academy of Sciences, P.O.Box 1410, 50-950, Wroclaw, Poland.

[4] International Laboratory for High Magnetic Fields and Low Temperatures, Gajowicka 95, 53-421 Wroclaw, Poland.

Abstract

The Andreev reflection spectra were studied for the first time in the magnetic superconductor $Dy_{0.6}Y_{0.4}Rh_{3.85}Ru_{0.15}B_4$. It is found that an external magnetic field (up to 0.7 $H_{c2}$) is a strong stimulator of superconductivity in contrast with traditional superconductors with a singlet pairing. The ratio $2\Delta/k_BT_c \approx 4.0$ obtained for some contacts is higher than the value 3.52 which is typical of conventional superconductors with a weak electron–phonon interaction. It has been proposed that in $Dy_{0.6}Y_{0.4}Rh_{3.85}Ru_{0.15}B_4$ a triplet mechanism of superconducting pairing is realized.

PACS numbers: 71.20.Eh, 74.20.Rp, 74.45.+c, 74.70.Dd

## 1. Introduction.

Several decades ago a high-temperature superconductivity was revealed in cuprates. The behavior of superconducting parameters in those materials could not be explained in the context of traditional BCS model. Later on it was shown that in this system a singlet mechanism of pairing exists, with a high degree of probability, with the total spin moment of a pair $S = 0$ and the orbit one $L = 2$ (the so-called $d$-wave state). It should be emphasized that in traditional singlet superconductors $S = 0$ and $L = 0$. At about the same time heavy–fermion superconductors UPt$_3$ [1] and Sr$_2$RuO$_4$ [2] were discovered where one could observe a still more exotic mechanism of superconducting pairing, namely, a triplet p-wave one with $S = 1$ and $L = 1$, as evidenced by direct experiments. Since for the triplet pairing the spins of paired electrons are unidirectional, this configuration is more stable against the effect of magnetic field than the conventional singlet one. Moreover, in compounds with the triplet pairing the superconductivity coexists very often with the magneto-ordered state of the lattice with nonzero magnetic moment. In this case the electrons are considered to be bound in a pair thanks to magnetic excitations and not to phonons as it takes place in traditional superconductors. The transition to a superconducting state in triplet superconductors occurs normally at rather low temperatures (e.g., $T_c \approx 0.7$ K in Sr$_2$RuO$_4$), making the problem of studying superconducting parameters more complicated.

We discovered recently a number of compounds Dy$_{1-x}$Y$_x$Rh$_4$B$_4$ (0≤x≤1) in which the superconductivity coexists with a ferrimagnetic ordering in a wide temperature range [3]. It has been shown that the behavior of superconducting parameters in Dy$_{0.8}$Y$_{0.2}$Rh$_4$B$_4$ (the gap $\Delta$ and second critical field $H_{c2}$) is nontrivial (they increase in some temperature and magnetic field ranges) [4, 5]. This fact as well as the relation $2\Delta/k_B T_c \approx 4.0$ (not 3.52 as in conventional superconductors) sent us to the viewpoint that the superconducting pairing in the compound under consideration is nontraditional. The behavior of $\Delta$ and $H_{c2}$ with increasing magnetic field suggested that the pairing mechanism could be a triplet one. It should be mentioned that in Dy$_{0.8}$Y$_{0.2}$Rh$_4$B$_4$ $T_c \approx 6$ K, i.e., it is by an order of magnitude higher than in Sr$_2$RuO$_4$. This makes such compounds more suitable for studying superconductivity and might help to elucidate the mechanism of pairing in the system discussed.

The paper concerns the study of magnetic field effect on the point contact (PC) Andreev reflection spectra dI/dV(V) of a further representative of the family of rare-earth rhodium borides, namely, Dy$_{0.6}$Y$_{0.4}$Rh$_{3.85}$Ru$_{0.15}$B$_4$ with a still higher superconducting transition temperature $T_c \approx 7$ K.

## 2. Experimental Procedure

The specimens $Dy_{0.6}Y_{0.4}Rh_{3.85}Ru_{0.15}B_4$ were prepared by using an argon arc melting of the initial components followed by its annealing for several days. The data of X-ray phase and X-ray diffraction analyses showed that the polycrystalline samples of a $LuRu_4B_4$-type crystal structure (space group I4/mmm) were single-phase.

The study concerned point contact Andreev-reflection spectra, *dI/dV(V)* characteristics, N-S contacts in a wide voltage bias range, much higher that the gap values, what permitted us to control the behavior of excess (Andreev) current and to eliminate unstable contacts from consideration. As a counter-electrode, it was used a mechanically sharpened and chemically etched thin Au wire ~ 0.1mm in diameter.

The measurements were performed at $T = 1.6$ K in magnetic fields varied from zero to critical values. Besides, several temperature sets of spectra were measured in the range from 1.6 K to critical point to determine the critical temperature of superconducting transition onset.

The PC spectra *dI/dV(V)* were taken by using the standard *lock-in* method and synchronous detection with a simultaneous computer recording  The spectra processing was carried out on the basis of the enhanced Blonder-Tinkham-Klapwiyk (BTK) theory [6-8] widely used for parameterization of point N-S contacts.

## 3. Experimental Results

The resistance transition in a superconducting state of $Dy_{0.6}Y_{0.4}Rh_{3.85}Ru_{0.15}B_4$ is shown in Fig.1. The estimation of superconducting transition temperature gives $T_c \approx 7$ K.

Figure 2 shows the PC spectra *dI/dV(V)* for the Au–$Dy_{0.6}Y_{0.4}Rh_{3.85}Ru_{0.15}B_4$ contact ($R_n \approx 3.7$ Ω) recorded in different magnetic field (0 ÷ $H_{c2}$) at $T = 1.6$ K. Similar magnetic field spectra were registered indiscriminately on stable contacts on which it had been possible to carry out a complete cycle of measurements in a temperature range between 1.6 and 2.0 K. A total number of such contacts was about ten with $R_n$ amounting to ~20 Ω. The superconducting transition onset temperature $T_c^{on}$ determined by the beginning of a distinct zero-bias maximum in the *dI/dV(V)* curve proved to be in the range from 6.7 to 6.9 K that is close to the value measured on the bulk sample (Fig.1). This provides support for the high quality of the contacts studied. The high contact quality is also evidenced by a considerable value of excess (Andreev) current $I_{exc}$, which little change in the over-gap voltage region ($V >> \Delta/e$), up to 80 % of the theoretically expected one.

Analysis of the spectral characteristics $dI/dV(V)$ at $T = 1.6$ K in different magnetic fields shows that they have two essential distinctions from those for traditional superconductors. First, in zero magnetic field the spectra in the neighborhood of $V = 0$ have no double gap maxima which are observed in conventional superconductors and usually are seen for the misfit of Fermi momenta in contacting electrodes or with the appearance of a thin dielectric layer on the N-S boundary [9]. Second, the difference of our spectra measured near 1.6 K (Fig.2) from the classical type is in the enhancement of the gap structure with increasing magnetic field. Beginning with a certain field value, there appear double maxima in the spectra as in N-S contacts of traditional superconductors. As the field continues to be increased further, the maxima intensity increases to a definite level and then the process proceeds in the inverse direction until the maxima are almost completely suppressed. At the same time the gap maximum voltages increase until they approach a definite value that remains constant up to $H_{c2}$. Thus, one can observe the clear evidence of increasing superconductivity by external magnetic field.

The magnetic field dependence of superconducting order parameter $\Delta(H)$ plotted at $T = 1.6$ K is shown in Fig.3. The parameter $\Delta$ was determined by fitting the modified BTK theory [7] to the experimental spectra (Fig.2). For comparison two theoretical dependences $\Delta(H)$ are shown in Fig.3 as well which are calculated for conventional second kind bulk superconductors [10] and for a thin film in a parallel magnetic field [11]. These dependences can be observed in traditional superconductors depending on relative orientation of contact axis and magnetic field.

The gap value for a number of contacts measured in zero magnetic field at $T \sim 1.6$ K proved to be up to 1.2 meV ($2\Delta/k_B T_c = 4.0$). It should be noted that $2\Delta/k_B T_c = 4.0$ is greater than the value for conventional superconductors ($2\Delta/k_B T_c = 3.52$), suggesting that either the electron-phonon interaction is strong (e.g., as in Hg and Pb) or the pairing mechanism differs from the traditional one.

From our viewpoint, the above-mentioned anomalous behavior of the PC spectra in magnetic field is responsible for by the fact that in the compound under consideration there occurs a triplet Cooper pairing, i.e. the total spin moment of a superconducting pair is $S = 1$. In such a case a moderate magnetic field is supposed to be not only far from breaking the superconducting state but in some cases to contribute to its stabilization or even enhancement [12].

If the effect of superconductivity stimulation was observed only in a narrow magnetic field range, it could be explained in the context of singlet pairing model as well. In such a case the non-collinear arrangement of magnetic momenta with total magnetization $M \neq 0$ would become a collinear one in a magnetic field with $M = 0$ and therefore, the internal factor suppressing the superconducting state dissappears. Some other possible causes of stimulation such as the occurrence of extraneous inclusions (different in phase composition) or dielectric layers into the point-contact area should not be taken into consideration because the critical parameters of all the contacts studied remained almost unchanged. Moreover, in some cases the excess current $I_{exc}$ reached ~ 80% of a theoretical probable value within the BTK model.

It is evident that to elucidate the nature of the anomalies observed in the PC spectra of $Dy_{0.6}Y_{0.4}Rh_{3.85}Ru_{0.15}B_4$ in magnetic fields and hence, the possibility of triplet pairing existence in the compound requires more comprehensive investigation by different methods.

**4. Conclusion.**

1. The PC Andreev reflection spectra $dI/dV(V)$ in the N-S contacts based on the magnetic superconductor $Dy_{0.6}Y_{0.4}Rh_{3.85}Ru_{0.15}B_4$ with $T_c^{on} = 6.7 \div 6.9$ K were studied for the first time in different magnetic fields.

2. It is found that as the magnetic field is increased the gap peculiarities in the spectra (and hence the gap/order parameter) are not only far from being shifted towards lower energies as it is supposed to be in the classical case, but, on the contrary, they move in the inverse direction enhancing their intensity. Upon reaching maximum and subsequent decreasing intensity, their position in the energy axis remains unchanged up to the critical magnetic field $H_{c2}$ at which the superconducting state disappears sharply.

3. It is shown that in some contacts $2\Delta/k_BT_c = 4$ what is greater than the value 3.52 expected for classical superconductors.

4. It is supposed that a triplet mechanism of superconducting pairing is realized in $Dy_{0.6}Y_{0.4}Rh_{3.85}Ru_{0.15}B_4$.

**Acknowledgement**

The work was partially supported by grants 12-02-01193 from Russian Fundamental Research Fund.

**Captions**

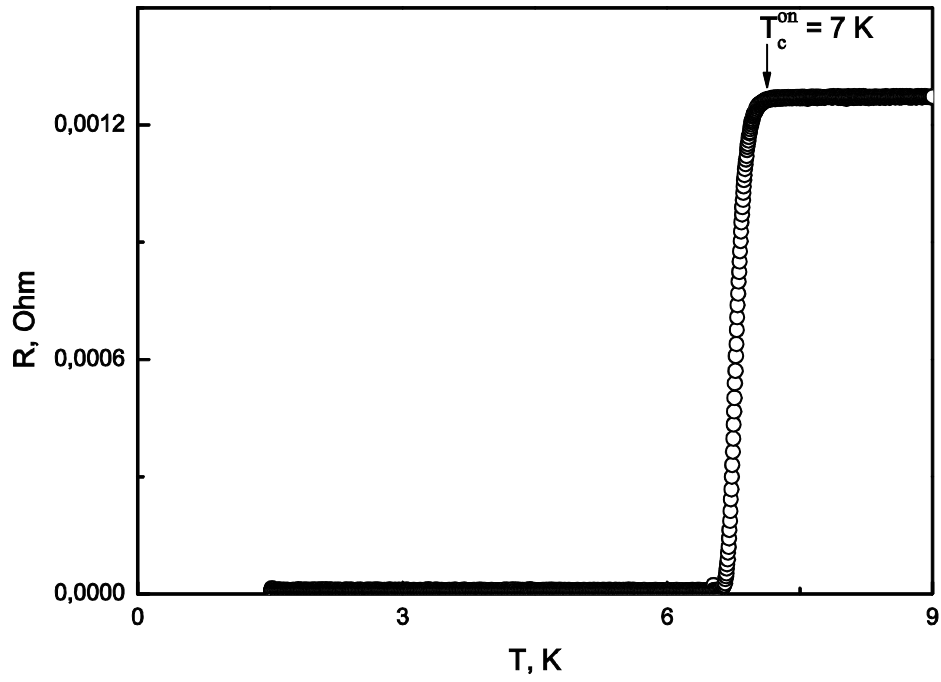

**Fig.1.** The resistive transition of the $Dy_{0.6}Y_{0.4}Rh_{3.85}Ru_{0.15}B_4$ sample into the superconducting state.

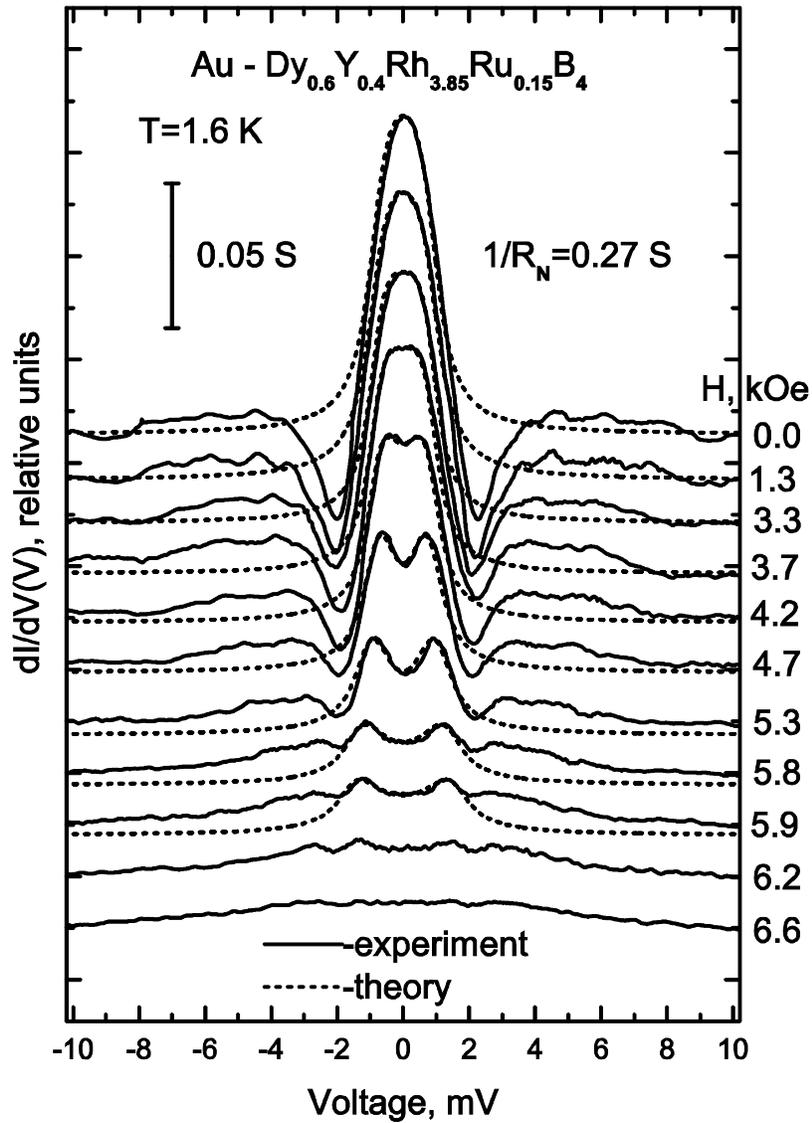

**Fig.2.** The representative set of Andreev spectra (*dI/dV(V)*) for a typical contact exhibiting a considerable enhancement of the gap structure in a magnetic field at $T = 1.6$ K. The BTK fitting of the spectra is shown by dash curves. The magnetic field is specified at each curves. For clearness, the curves are arbitrarily displaced vertically.

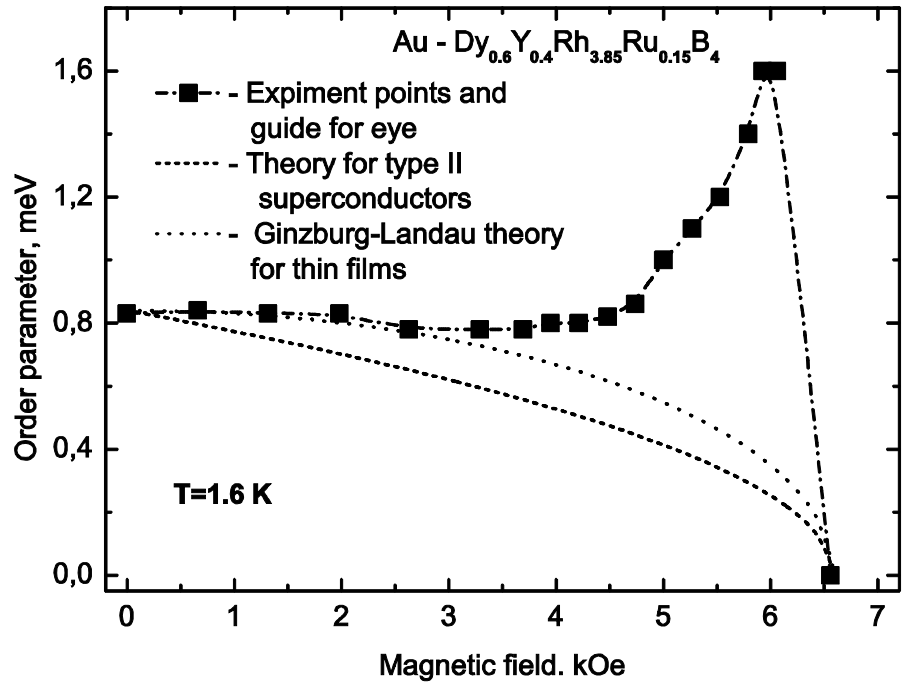

**Fig.3.** The dependence of the order parameter upon the magnetic field Δ*(H)* at *T* ≈ 1.6 K for the contact whose spectra are illustrated in Fig.3. For comparison, two theoretical dependences (broken lines) are shown, which are possible in contacts based on ordinary superconductors when the contact axis is along or perpendicular to the field.